\documentclass[superscriptaddress,twocolumn,amsmath,amssymb,showpacs,floatfix,prb]{revtex4}

\usepackage{graphicx}
\usepackage{dcolumn}
\usepackage{bm}
\usepackage{rotating}
\usepackage{booktabs}
\usepackage{color}

\newcommand{\CFO}{CuFeO$_2$}

\newcommand{\tna}{$T_{\text{N}1}$}
\newcommand{\tnb}{$T_{\text{N}2}$}

\newcommand{\Bpar}{$B^{\parallel}$}
\newcommand{\Bpara}{$B^{\parallel}_{c1}$}

\newcommand{\Bparb}{$B^{\parallel}_{c2}$}

\newcommand{\Bparc}{$B^{\parallel}_{c3}$}

\newcommand{\Bpard}{$B^{\parallel}_{c4}$}
\newcommand{\Bpare}{$B^{\parallel}_{c5}$}

\newcommand{\Bperpa}{$B^{\perp}_{c1}$}

\newcommand{\Bperpb}{$B^{\perp}_{c2}$}
\newcommand{\Bperpc}{$B^{\perp}_{c3}$}

\newcommand{\figwidth}{8.5cm}

\begin{document}

\title{High-field recovery of the undistorted triangular lattice in the frustrated metamagnet \CFO.}

\author{T.T.A. Lummen}
\affiliation{Zernike Institute for Advanced Materials, University of
Groningen, Nijenborgh 4, 9747 AG Groningen, The Netherlands}

\author{C. Strohm}
\altaffiliation{Current address: European Synchrotron Radiation
Facility (ESRF) - P.O. Box 220, 38043 Grenoble, France}
\affiliation{Institut N\'{e}el, CNRS et Universit\'{e} Joseph
Fourier, BP 166, F-38042, Grenoble Cedex 9, France}

\author{H. Rakoto}
\affiliation{Laboratoire National des Champs Magn\'{e}tiques
Puls\'{e}s (LNCMP), 143 avenue de Rangueil, 31400 Toulouse, France}

\author{A.A. Nugroho}
\affiliation{Departemen Fisika FMIPA, Institut Teknologi Bandung,
Jl. Ganesa 10, Bandung 40132, Indonesia}

%\author{I.P. Handayani}
%\affiliation{Zernike Institute for Advanced Materials, University of
%Groningen, Nijenborgh 4, 9747 AG Groningen, The Netherlands}

%\author{G. Dhalenne}
%\affiliation{Laboratoire de Physico-Chimie de l'Etat Solide, CNRS,
%UMR8182, Universit\'{e} Paris-Sud, B\^{a}timent 414, 91405 Orsay,
%France}

%\author{A. Revcolevschi}
%\affiliation{Laboratoire de Physico-Chimie de l'Etat Solide, CNRS,
%UMR8182, Universit\'{e} Paris-Sud, B\^{a}timent 414, 91405 Orsay,
%France}

\author{P.H.M. van Loosdrecht}
\email{P.H.M.van.Loosdrecht@rug.nl} \affiliation{Zernike Institute
for Advanced Materials, University of Groningen, Nijenborgh 4, 9747
AG Groningen, The Netherlands}

\date{\today}

\begin{abstract}
Pulsed field magnetization experiments extend the typical
metamagnetic staircase of \CFO\ up to 58 T to reveal an additional
first order phase transition at high field for both the parallel and
perpendicular field configuration. Virtually complete isotropic
behavior is retrieved only above this transition, indicating the
high-field recovery of the undistorted triangular lattice. A
consistent phenomenological rationalization for the field dependence
and metamagnetism crossover of the system is provided, demonstrating
the importance of both spin-phonon coupling and a small
field-dependent easy-axis anisotropy in accurately describing the
magnetization process of \CFO.
\end{abstract}


\maketitle

\begin{figure}[htb]
\centering
\includegraphics[width=\figwidth]{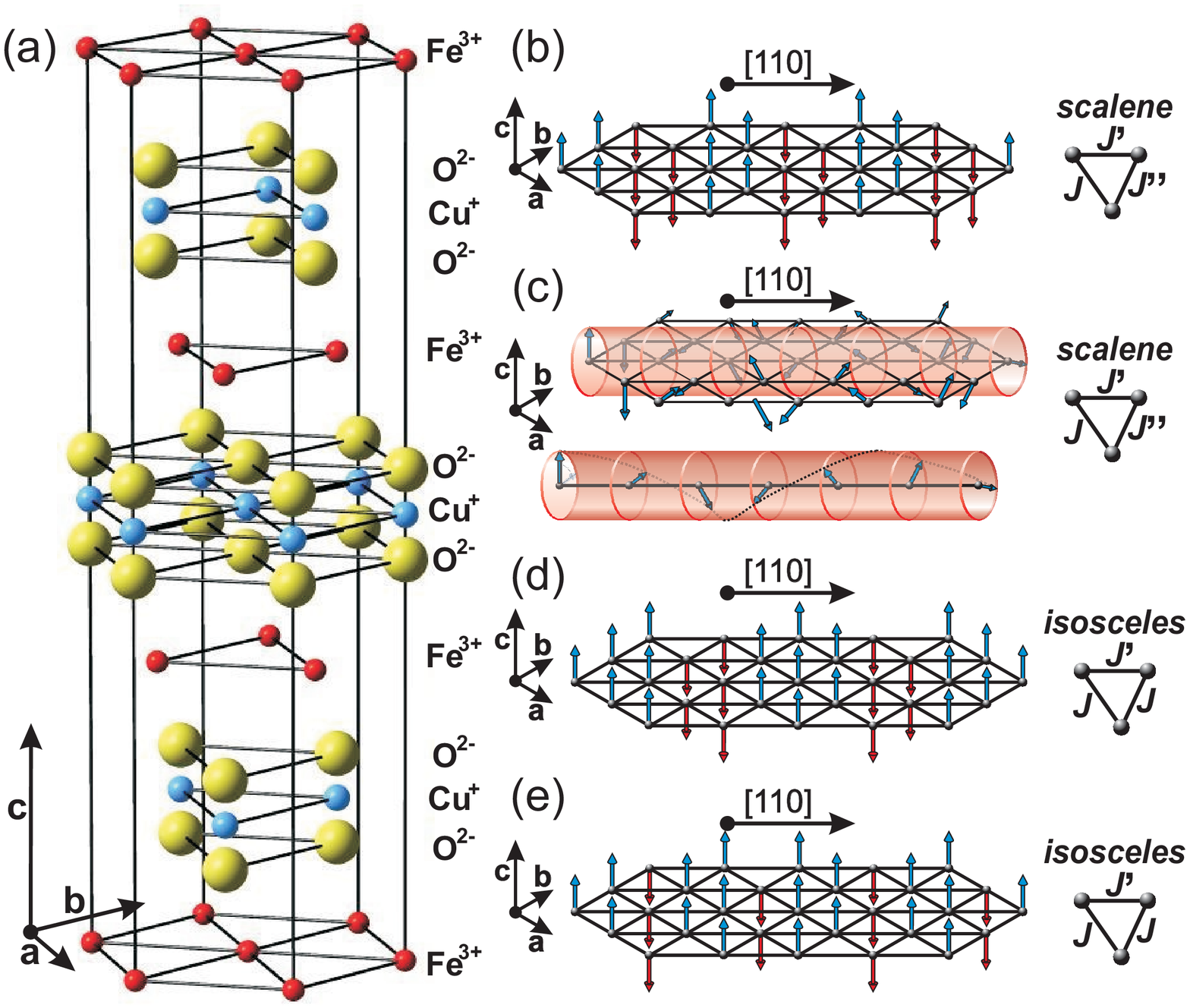}
\caption{\label{fig1} \small (Color online) \textbf{(a)} Schematic
crystal structure of \CFO\ (\textit{R\={3}m},
\textit{a}=\textit{b}=3.03\AA, \textit{c}=17.17\AA). To avoid
confusion, crystal directions are referred to using the hexagonal
description throughout the article. \textbf{(b-e)} Spin structures
and lattice symmetries in various field-induced phases of \CFO\
($B$$\parallel$$c$): collinear four-sublattice (4SL) phase
(\textbf{(b)}), ferroelectric helical phase (\textbf{(c)}),
collinear 5SL phase (\textbf{(d)}) and collinear 3SL phase
(\textbf{(e)}).}
\end{figure}

Metamagnetism typically refers to any material that, upon variation
in the externally applied magnetic field, exhibits an abrupt change
in magnetization. In general, the phase diagrams of materials
undergoing field-induced magnetic transitions can be rationalized
according to the degree of magnetic anisotropy in the materials
\cite{stryjewski77}. In highly anisotropic systems, spins are
effectively restricted to align (anti-)parallel to the magnetic
easy-axis and magnetic transitions typically involve discontinuous
spin reversals, leading to first-order type metamagnetic
transitions. As for isotropic (weakly anisotropic) systems this
directional restriction is relieved (strongly reduced), transitions
in such materials often reflect the onset of a continuous, second
order type reorientation of the local spins. Another source of
exotic magnetic transitions is geometrical magnetic frustration,
which occurs when a specific lattice geometry prevents the
simultaneous minimization of all magnetic exchange interactions,
thus introducing a high spin degeneracy \cite{ramirez94}. The
simultaneous occurrence of both these phenomena and the interplay
between them leads to intricate, diverse and rich physics, yielding
many captivating magnetic phases ranging from spin liquids and ices
to multiferroic spiral phases
\cite{penc04,greedan06,kimura06,ueda06}.\newline \indent Here the
focus is on the \textit{delafossite} semiconductor \CFO, an
archetype triangular lattice antiferromagnet, in which the Fe$^{3+}$
ions stack in hexagonal layers along the c-axis (Fig.
\ref{fig1}(a)). In spite of the expected Heisenberg nature of the
Fe$^{3+}$ spins ($3d^5$, $S = 5/2$, $L = 0$), \CFO\ does not order
in the noncollinear $120^{\circ}$ spin configuration at low
temperature. Instead, after undergoing successive phase transitions
at \tna\ $\approx 14$ K and \tnb\ $\approx 11$ K, lowering the
symmetry from hexagonal (\textit{R\={3}m}) to monoclinic
\cite{terada06_1,ye06,terada06_2}, the system adopts a collinear,
two-up two-down order, with moments aligned (anti)parallel along the
$c$-axis (Fig. \ref{fig1}(b)) \cite{mekata92}. The collinear ground
state is supposedly stabilized through the strong spin-lattice
coupling in \CFO\ \cite{ye06,terada06_1,terada06_2,wang08}, which
induces a structural distortion through the 'spin Jahn-Teller'
effect \cite{yamashita00,tchernyshyov02}. Alternatively, this
scalene triangle distortion has been argued to induce an easy axis
anisotropy, which was also used to account for the the observed
Ising-like magnetism \cite{ye07,petrenko05,terada07_3}. An
intriguing behavior arises when \CFO\ is subjected to an external
magnetic field $B$ below \tnb. With $B$$\parallel$$c$, the spin
system has been found to successively assume a proper helical
ordered ferroelectric phase, a collinear three-up two-down ordered
phase, a phase with a magnetization plateau at one-third of the
saturation value and a phase with steadily increasing magnetization
\cite{mitsuda00_1,petrenko05,terada06_3,kimura06,terada07_3}.
Particular attention has gone to the ferroelectric helical ordered
phase \cite{kimura06,arima07,nakajima07,nakajima08}, which has since
also been stabilized in zero field through Al$^{3+}$ or Ga$^{3+}$
substitution \cite{seki07,kanetsuki07,terada08}. With increasing
$B$$\perp$$c$, the magnetization has been found to first increase
steadily, then halt at a one-third plateau before resuming a
(\textit{quasi}-)linear increase at higher fields
\cite{petrenko05,terada07_3}. This article presents pulsed field
magnetization experiments, which extend the metamagnetic staircase
of \CFO\ up to fields exceeding 58 T \cite{samples}. An additional
high field first order phase transition is observed for both
parallel and perpendicular configurations, above which virtually
complete isotropic behavior is retrieved, indicating the recovery of
the undistorted triangular structure. Moreover, a consistent
phenomenological interpretation is provided, combining all magnetic
terms deemed to be important in \CFO.\newline
\begin{figure}[htb]
\centering
\includegraphics[width=\figwidth]{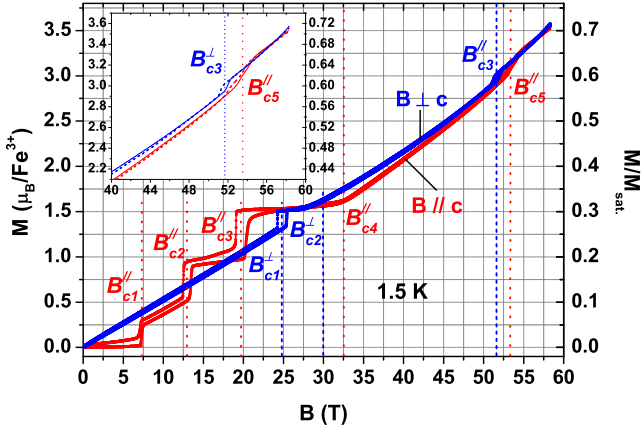}
\caption{\label{fig2} \small (Color online) Pulsed field
magnetization process for both $B$$\parallel$$c$ (red/light gray)
and $B$$\perp$$c$ (blue/dark grey) at 1.5 K. Inset: Zoom-in on high
field region.}
\end{figure}
\indent As depicted in Fig. \ref{fig2}, the magnetization process
for both $B$$\parallel$$c$ and $B$$\perp$$c$ shows a cascade of
phase transitions, in excellent agreement with literature
\cite{ajiro94,mitsuda00_1,petrenko05,terada06_3,kimura06,terada07_3}.
As \Bpar\ increases, the spin system successively rearranges to the
helical ordered phase at \Bpara\ $\simeq$ 7.2 T (Fig. \ref{fig1}(c))
and the collinear three-up two-down phase at \Bparb\ $\simeq$ 13.0 T
(Fig. \ref{fig1}(d)). Recent synchrotron x-ray-diffraction studies
up to 40 T revealed the strong correlation between the spin
Jahn-Teller lattice distortion and the magnetization process in
\CFO; as $M$ increases, the extent of the distortion decreases
accordingly \cite{terada06_3,terada07_3}. Since the induced magnetic
anisotropy ($D$) is directly coupled to the distortion, one may
assume it also diminishes correspondingly with $B$; exhibiting steps
at first order transitions and continuously decreasing in
(\textit{quasi-})linear phases. Moreover, at \Bparb, the symmetry of
the distortion is increased, yielding a lattice of isosceles
triangles \cite{terada06_2}. At \Bparc\ $\simeq$ 19.7 T, the
magnetization jumps to a plateau phase at one-third of saturation,
signaling a collinear two-up one-down order (Fig. \ref{fig1}(e))
\cite{terada06_3,terada07_3}, which is consistent with an expected
nonzero $D$ and pulsed field nuclear forward scattering experiments
\cite{strohm}. At \Bpard\ $\simeq$ 32.4 T, the system undergoes a
second order transition, above which $M$ starts growing
continuously, indicating gradual canting of the spins away from
collinearity. By extending the experimental field of view to higher
fields, the persistence of this (\textit{quasi-})linear increase up
to \Bpare\ $\simeq$ 53.3 T could be determined, where an additional
metamagnetic transition is identified. Above this transition, up to
58.3 T, $M$ grows steadily once more, with slightly different slope.
In short, below \Bpard\ magnetic transitions are of first order,
exhibiting significant hysteresis and large magnetization steps.
From \Bpard\ on (second order transition), $M$ mostly increases
continuously with \Bpar, and magnetization plateaus are absent. In
terms of metamagnetism, this corresponds to a crossover from a
highly anisotropic regime (with abrupt spin flips) to a weakly
anisotropic regime (continuous spin reorientation), which is in line
with the notion of progressive symmetry increase and thus magnetic
anisotropy reduction upon increasing $B$. \newline \indent For
$B$$\perp$$c$, the magnetization process is quite different.
Starting from the zero-field collinear two-up two-down phase, $M$
shows a steady increase up to \Bperpa\ $\simeq$ 24.8 T, where a
first order transition brings the system in a plateau phase at
one-third of saturation, implying a three sublattice (3SL)
structure. Above \Bperpb\ $\simeq$ 30.0 T, the system exhibits a
steady increase of $M$ after undergoing a second order phase
transition, indicating a continuous spin reorientation. The data in
Fig. \ref{fig2} show this behavior persists up to \Bperpc\ $\simeq$
51.6 T where the system undergoes an additional first order
transition, similar to that at \Bpare\ for $B$$\parallel$$c$.
Contrary to previous claims \cite{terada07_3}, the behavior clearly
remains anisotropic up to these transitions. Though left
unaddressed, a corresponding feature can also be observed around 52
T in the ($dM/dB$) vs.$\;B$ data previously recorded in a single
turn coil measurement up to 100 T (8 K) \cite{ajiro94}. At fields
above both these transitions, $M$ shows virtually isotropic
behavior, growing (\textit{quasi-})linearly up to 58.3 T. This
absence of anisotropy suggests a full symmetry recovery and thus
retrieval of the undistorted triangular lattice at these
fields.\newline \indent To elucidate the nature of the high field
spin structures, we introduce a simple classical spin model for a
single triangular sheet, which includes the primary terms in the
spin Hamiltonian:
\begin{eqnarray}
H = & -g\mu \mbox{\boldmath{$B$}}\cdot\displaystyle
\sum_{i}\mbox{\boldmath{$S$}}_{i} + \displaystyle \sum_{i,j}
J_{ij}\mbox{\boldmath{$S$}}_{i}\cdot
\mbox{\boldmath{$S$}}_{j} \nonumber\\
&  - \displaystyle \sum_{\langle i,j \rangle}
bJ_{ij}(\mbox{\boldmath{$S$}}_{i}\cdot
\mbox{\boldmath{$S$}}_{j})^{2} - D(B)\displaystyle \sum_{i}
S_{iz}^2,\label{Hamiltonian}
\end{eqnarray}
where $J_{ij}$ is the exchange coupling between sites $i$ and $j$,
$b$ is an effective (nearest-neighbor only) biquadratic interaction
originating from the spin-lattice coupling (bond-phonon
model)\cite{penc04,bergman06} and $D(B)$ is the anisotropy constant
(\textgreater\ 0 for an easy-axis along $z$). The first (Zeeman) and
last (anisotropy) terms sum over all sites $i$ in the magnetic unit
cell, while the summation in the exchange and biquadratic terms
includes all interactions within that unit cell. The high field
magnetization process is qualitatively similar for both field
configurations; first the spin system exhibits a one-third
magnetization plateau, implying a three sublattice (3SL) structure,
after which $M$ starts increasing steadily, indicating a continuous
reorientation of the 3SL spins. To capture this magnetic behavior,
we thus study the spin Hamiltonian of the 3SL structure on a single
sheet. As the lattice distortion persists up to at least 40 T for
both field configurations \cite{terada07_3}, a nonzero $D$ may be
expected up to these fields. Having three inequivalent spins, there
are three unique first- ($J$), second- ($J^{\dag}$) and
third-neighbor ($J^{\ddag}$) couplings per spin in one magnetic unit
cell (Fig. \ref{fig3}(a)). The Hamiltonian becomes (using $g=2$,
$\mbox{\boldmath{$S$}}_{i}=\mbox{\boldmath{$e$}}_{i}S$ (unit vector
$\mbox{\boldmath{$e$}}$, classical spins) and
$p_{ij}=\mbox{\boldmath{$e$}}_{i}\cdot\mbox{\boldmath{$e$}}_{j}$):
\begin{eqnarray}
H = &-2\mu_{B} S\mbox{\boldmath{$B$}}\cdot\displaystyle \sum_{i}\mbox{\boldmath{$e$}}_{i}+{}CS^{2}(p_{12}+p_{13}+p_{23})+{}9J^{\dag}S^{2} \nonumber \\
&{}-{}GS^{4}(p_{12}^{2}+p_{13}^{2}+p_{23}^{2})-{}D(B)S^{2}\displaystyle
\sum_{i} e_{i,z}^2,\label{model}
\end{eqnarray}
where the exchange constant $C=(3J+3J^{\ddag})$ and the spin-lattice
constant $G=3bJ$. The anisotropy $D(B)$ is approximated to be
'\textit{anti}-proportional' to $M(B)$; as $M$ approaches
saturation, $D$ vanishes accordingly. Note that the inclusion of
further neighbor interactions has only a trivial effect; the third
neighbor interactions merely add to $C$, while the second neighbor
interactions only shift the total energy as a whole.\newline \indent
\begin{figure}[htb]
\centering
\includegraphics[width=\figwidth]{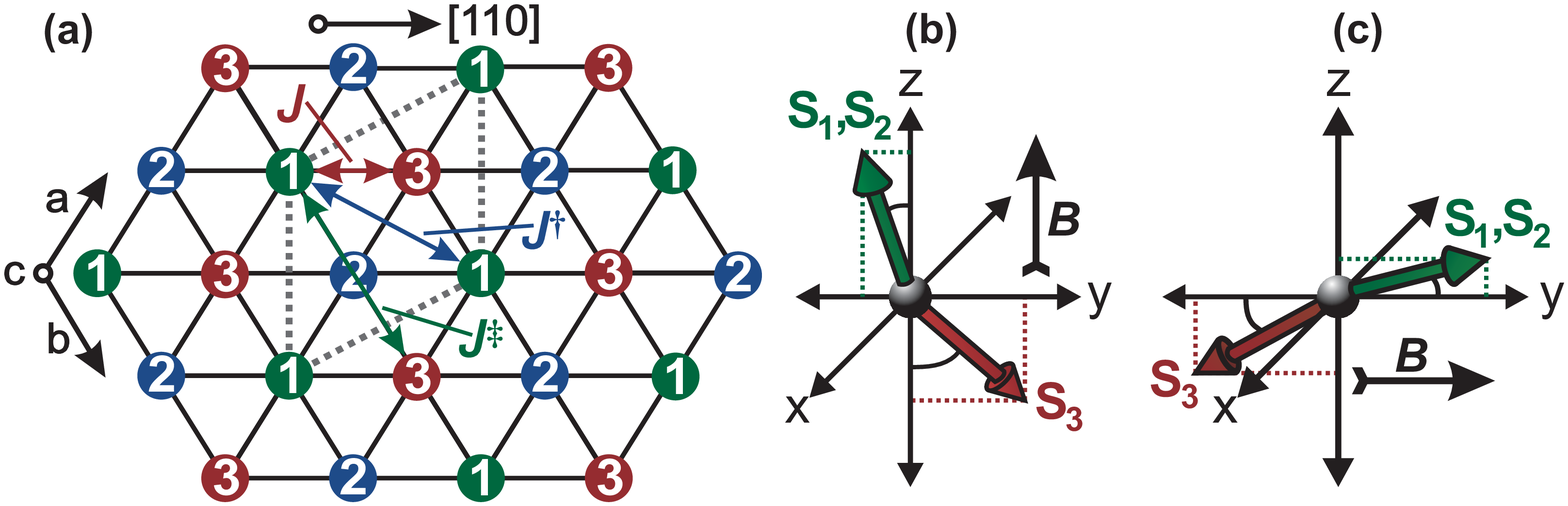}
\caption{\label{fig3} \small (Color online) \textbf{(a)} Exchange
interactions in the 3SL structure on the 2D isosceles triangle
lattice. Minimum energy solutions of eq. \ref{model} for a 3SL
structure with finite $D$ are depicted for $B$$\parallel$$c$
(\textbf{(b)}) and $B$$\perp$$c$ (\textbf{(c)}).}
\end{figure}
We determined the spin directions $\mbox{\boldmath{$e$}}_{1}$,
$\mbox{\boldmath{$e$}}_{2}$ and $\mbox{\boldmath{$e$}}_{3}$
corresponding to the minimum energy per unit cell at a given field
\mbox{\boldmath{$B$}} by performing numerical minimization of eq.
\ref{model}. In examining the corresponding simulated magnetization
curves, one finds that having both the anisotropy constant $D(B)$
and the spin-lattice constant $G$ nonzero is a prerequisite for good
qualitative agreement with experiment; with $D=0$ the magnetization
process becomes isotropic, while a nonzero $G$ is required to
stabilize a magnetization plateau for $B$$\perp$$c$. Matching the
simulated magnetization curves with experiment yields estimates for
the exchange, anisotropy and spin-lattice constants of $C$ = 1.32
meV, $D(25$T$)$ = 0.021 meV (plateau) and $G$ = 0.0074 meV,
respectively, the resulting curves are plotted in Fig. \ref{fig4}.
Taking only first-neighbor interactions, we can estimate $JS^{2}$ as
$\sim$ 2.76 meV (32.0 K) and $DS$ (at 25T) as $\sim$ 0.052 meV (0.6
K), in line with previous estimates \cite{wang08,petrenko05,ye07}.
For $GS^{4}$ we estimate $\sim$ 0.29 meV (3.4 K), yielding a
dimensionless biquadratic coupling $b$ of $\sim$ 0.0056 (compared to
$\sim$ 0.008 using estimates from ref. 11). With these parameters
the simulations are in striking agreement with experiment. The
spin-lattice interaction $G$ (non-directional) stabilizes the
one-third magnetization plateau in both configurations, while the
directional anisotropy interaction $D$ widens the plateau for
$B$$\parallel$$c$ and narrows it for $B$$\perp$$c$, leading to the
difference in plateau-widths and above-plateau increase of $M$.
Moreover, a nonzero $G$ also induces a positive
$\partial^{2}M/\partial B^{2}$ in the latter, as is observed in
experiment.\newline \indent
\begin{figure}[htb] \centering
\includegraphics[width=\figwidth]{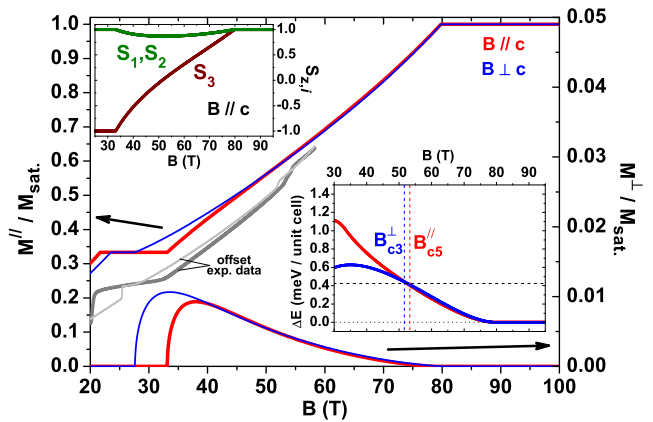}
\caption{\label{fig4} \small (Color online) Simulated magnetization
process (eq. \ref{model}) in \CFO\ for both $B$$\parallel$$c$
(thick, red) and $B$$\perp$$c$ (thin, blue). $M$ is plotted in the
both the field direction (left axis, $M^{\parallel}$) and the plane
perpendicular to $B$ (right axis, $M^{\perp}$). Thick dark (thin
light) grey lines depict offset experimental data for
$B$$\parallel$$c$ ($B$$\perp$$c$). Upper inset: field-dependence of
z-components of individual spins for $B$$\parallel$$c$. Lower inset:
estimated magnetic energy gain w.r.t. isotropic spin structure (see
text).}
\end{figure}
According to the minimum energy solution the three spins evolve as
follows for $B$$\parallel$$c$: below \Bpard, the system is in the
collinear 3SL state, with two spins ($S_{1}$ and $S_{2}$) parallel
to \Bpar, and one ($S_{3}$) antiparallel. As depicted in the upper
inset of Fig. \ref{fig4}, from \Bpard\ on, the 'down' spin starts
continuously tilting from 'down' to 'up', thereby increasing $M$. In
optimizing the overall magnetic energy, the two 'up' spins respond
by first moving slightly away from the c-axis in the opposite
direction, before gradually returning after the 'down' spin has
passed the basal plane. Due to the finite $D$ and $G$, the two 'up'
spins remain collinear (Fig. \ref{fig3}(b)) and the system also
acquires a small in-plane magnetization, which quickly grows and
slowly decreases with \Bpar\ above \Bpard\ (Fig. \ref{fig4}).
Although the process is qualitatively similar for $B$$\perp$$c$
(Fig. \ref{fig3}(c)), quantitatively it differs slightly due to the
orthogonality of the field- and anisotropy directions in this
case.\newline \indent The model does not directly account for the
additionally observed high field transitions; it predicts continuous
evolution toward full saturation. However, the model assumes a
distortion-induced finite anisotropy $D(B)$, while the associated
elastic energy cost is not included in the all-magnetic Hamiltonian.
The amount of magnetic energy the system gains upon having a
distortion (and thus finite $D$) can be approximated by taking the
difference between the energy of an isotropic spin configuration
(e.g. a \textit{canted} $120^{\circ}$ configuration, with all three
spins tilted away from the field-direction, while their projections
in the orthogonal plane keep mutual $120^{\circ}$ angles) and the
minimum energy solution of eq. \ref{model} (lower inset of Fig.
\ref{fig4}). The first order transitions at \Bpare\ and \Bperpc\ can
then be identified as the point where the magnetic energy gain no
longer outweighs the elastic cost of the distortion, upon which the
system reverts to the undistorted triangular lattice, which is
corroborated by the observed isotropy in $M$ above \Bpare\ (Fig.
\ref{fig2}). We note the estimated magnetic energy gain at this
point is $\sim$ 0.42 meV per unit cell (3 spins), which is
approximately the temperature scale of the experimental data ($3kT$
at 1.5 K is $\sim$ 0.39 meV). This is consistent with the fact that
\Bpare\ and \Bperpc\ are observed to shift toward lower fields with
increasing temperature. For an isotropic lattice above \Bpare, the
model predicts a degenerate set of spin structures, among which the
aforementioned \textit{canted} $120^{\circ}$ structure.\newline
\indent Despite its satisfactory and intuitive results, our simple
model has its limitations. Though the low field collinear phases can
be modeled using eq. \ref{Hamiltonian}, being a phenomenological
model meant to describe the high field phases, it does not capture
the complex helical ferroelectric phase. A full quantitative
description of \CFO\ would require the inclusion of finite
temperature, three dimensionality (recent work showed the
significance of interplane couplings
\cite{petrenko05,terada07_2,ye07,fishman08}), a more refined phonon
model, quantum spins and possibly other interactions
\cite{plumer08}.\newline \indent Concluding, through pulsed field
magnetization experiments, the metamagnetic staircase characteristic
of \CFO\ was extended to up to 58.3 T, revealing an additional first
order phase transition for both magnetic field configurations, which
is proposed to be due to a reversed spin Jahn Teller transition.
Above this transition, virtually complete isotropic behavior is
retrieved. A highly consistent phenomenological rationalization for
the magnetization process in both magnetic field configurations is
developed, combining for the first time all magnetic terms deemed of
importance in \CFO. Numerical simulations based on the corresponding
classical model prove the pertinence of both spin-lattice and
field-dependent anisotropy interactions in \CFO. Combined with the
magnetization measurements, a recovery of the undistorted triangular
lattice structure is anticipated at high fields. The underlying
intuitive concept of progressive symmetry increase as the degree of
frustration in spin Jahn-Teller distorted systems diminishes is
rather universal, as it relies solely on energy arguments. Indeed, a
similar high field transition has been observed recently in
HgCr$_{2}$O$_{4}$ and corresponding transitions may be expected in
related spinel systems \cite{ueda06}.\newline \indent The authors
would like to thank F. de Haan and D. Maillard for technical
support. Financial support from the Agence Nationale de Recherche
under contract NT05-4\_42463 is gratefully acknowledged.

%\bibliography{CuFeO2}

\begin{thebibliography}{31}
\bibitem[{sam()}]{samples}
\bibinfo{note}{Samples were prepared from an oriented single crystal grown
  using the floating zone technique.}
\end{thebibliography}

\end{document}